\documentclass[reprint, amsmath,amssymb,aps, prb,superscriptaddress]{revtex4-1}
\usepackage{amsfonts,amssymb}
\usepackage{graphicx}
\usepackage{hyperref}    
\usepackage{bm}          
\usepackage{natbib}
\usepackage{soul} 
\usepackage{color}
\usepackage{longtable}
\usepackage{ textcomp }
\DeclareMathOperator{\arccot}{arccot}

\begin{document}
\author{K. V. Reich}
\email{kreich@umn.edu}
\affiliation{Fine Theoretical Physics Institute, University of Minnesota, Minneapolis, MN 55455, USA}
\affiliation{Ioffe Institute, St Petersburg, 194021, Russia}
\author{B. I. Shklovskii}
\affiliation{Fine Theoretical Physics Institute, University of Minnesota, Minneapolis, MN 55455, USA}

\title{Exciton Transfer in Array of Epitaxially Connected Nanocrystals}

\begin{abstract}
  Recently, epitaxially connected at facets semiconductor nanocrystals (NCs) have been introduced to fascilitate the electron transport between nanocrystals. To fully deploy their potential a better understanding of the exciton transfer  between connected NCs is needed. We go beyond the two well-known transfer mechanisms suggested by F{\"o}rster and Dexter and  propose a third mechanism of exciton tandem tunneling. The tandem tunnelling occurs through the intermediate state in which electron and hole are in different NCs.
  The corresponding rate for exciton hops  is larger than  the Dexter rate and for Si is even much larger that the  F{\"o}rster one.
\end{abstract}

\maketitle

\begin{figure}[h]
  \includegraphics[width=0.45\textwidth]{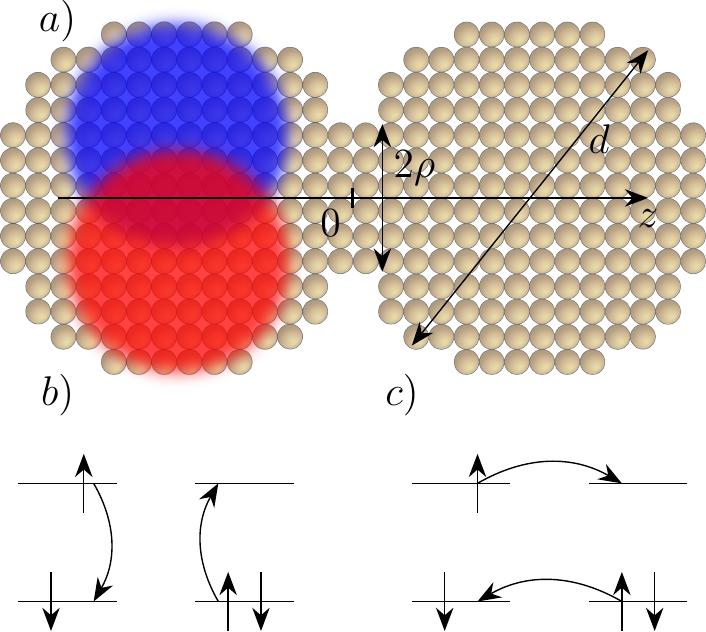}
  \caption{a) The exciton transfer between  spherical  NCs  with diameter $d$, which are epitaxially connected at the disk-like facet with the radius $\rho$.  The exciton is shown as  electron (blue) and hole (red) clouds. b) In the F{\"o}rster mechanism the electron in the conduction band of the left NC recombines with the hole in the valence band of the left NC  exciting the exciton in  the right NC \textit{via} dipole-dipole interaction. c) The  exciton tunneling from the left NC to the right one. In Dexter mechanism this happens by \textit{via} single exchange process of two electrons, while the tandem tunneling occurs in the second order perturbation theory by two one-electron hops through the intermediate state where electrons are in the same NC with large Coulomb energy.}\label{fig:Scheme}
 \end{figure}

Semiconductor nanocrystals (NCs) have shown great potential in optoelectronics applications such as solar cells \cite{ki_Supran_Bawendi_Bulovic_2012,Kortshagen_Si_solar_cell,gur_air-stable_2005}, light-emitting diodes \cite{LED_Yang_2015,Near_infrared_LED_2016,Robel_Lee_Pietryga_Klimov_2013}, field-effect transistors \cite{Kagan_surface_trap_passivation,Murray_Bandlike,Reich_Shklovskii,iu_Crawford_Hemminger_Law_2013,Kagan_FET_2016} and mid-infrared detectors \cite{Philippe_HgTe,Philippe_mid_infrared_2016} by virtue of their size-tunable optical and electrical properties and low-cost solution-based processing techniques \cite{Lee_Kovalenko_Shevchenko_2010}. As grown NCs are covered by ligands which deplete conductivity of NC arrays. For applications one needs a good electronic transport in a NC array. Substantial improvement of the transport properties of a NC array was achieved  by replacing of long ligands with shorter ones  \cite{Kagan_QD_review,Murray_Bandlike,iu_Crawford_Hemminger_Law_2013,Talapin_2013_high_mobility}. Recent progress \cite{Thimsen_ZnO_MIT,Kagan_ALD_NC_2014,Review_NC__touching_array,strong_coupling_superlattice,Review_NC__touching_array,QD_epitaxial_conneccted,lattice_NC_Tobias, facet_PbSE_mobility_2,facet_PbSE_mobility_1} lead to the creation of NCs which touch each other by facets or are epitaxially connected  and as a result demonstrate good conductivity \cite{transport_lattice_QD,lattice_NC_Vanmaekelbergh_THzmobility,Talapin_NCs_bridges}.
Fig. \ref{fig:Scheme}a shows an example of two NCs epitaxially connected at a facet with a small contact radius $\rho$ NCs.
Electron tunneling through a small facet leading to variable range hopping of electrons in doped NC array was studied theoretically \cite{localization_length_NC} and the criterium of the insulator-metal transition  was derived \cite{Ting_MIT}.
The transition was approached \textit{via} doping of NCs \cite{Ting_MIT}  or crossed \textit{via} increasing  contact radius \cite{Thimsen_ZnO_MIT}.

In optical devices based on NC arrays  absorption of a light quantum results in  the creation of an exciton (a bound electron-hole pair) in a NC.
An exciton can hop between nearest neighbour NCs.
Corresponding  diffusion coefficient and diffusion length were studied experimentally \cite{Kagan_Foster_CdSe,Klimov_exciton_transfer,Kinoshita_exciton_hopping,Klimov_Foster,Zamkov_exciton_diffusion,Tisdale_exciton_migration,uben_Beard_Luther_Johnson_2013}.
The diffusion length of excitons  sets the volume from which the light energy is harvested in solar cells .
Thus, the mechanism of exciton transfer between NCs is central to a NC device design.
This paper addresses exciton hopping between nearest epitaxially connected NCs.

In a typical  array NCs have slightly different diameters and therefore excitons have different ground state energies.
At low temperatures an exciton hops from a small NC to a larger one, where its energy is smaller.
This leads to a red shift of photoluminscence \cite{Kagan_Foster_CdSe}.
On the other hand, at larger temperatures an exciton can hop \cite{Zamkov_exciton_diffusion} even from a larger NC to a smaller one with absorption of a phonon.
In both cases the exciton transfer rate $1/\tau$  between two nearest-neighbor NCs  is determined by the Fermi golden rule and is proportional to the square of  the matrix element for the exciton transfer $M$.
($2M$ is the energy splitting between symmetric and antisymmetric states of  two resonance nearest neighbor NCs.)
In this work we concentrate on this matrix element.
It is known that an  exciton can  hop from one NC to another \textit{via} dipole-dipole F{\"o}rster mechanism \cite{Kagan_Foster_CdSe,Tisdale_Foster_radius,Tisdale_subdiffusive_transport} (see Fig. \ref{fig:Scheme}b).
This mechanism does not require tunneling of an electron or a hole between NCs and, therefore,  dominates when NCs are separated by long ligands.
Epitaxial connection of  NCs \textit{via} small facet with radius $\rho$ (see Fig.~\ref{fig:Scheme}a) does not change the F{\"o}rster rate.
On the other hand, tunneling of electrons and holes between epitaxially connected NCs opens additional possibilities for the exciton transfer.

 To describe them we assume that  a NC is almost a sphere of diameter  $d \simeq 3-8 ~\mathrm{nm}$.
 In an isolated NC the electron and hole wave functions vanish at the NC surface, due to a large confining potential  barriers created by the insulator matrix surrounding each NC.
 Under these conditions the ground state kinetic energy of an electron and hole is
 \begin{equation}
  \label{eq:Delta}
\Delta_{e,h} = 2\pi^{2} \frac{\hbar^2}{m_{e,h}d^2},  
\end{equation}
where $m_{e,h}$ are the effective masses of an electron and hole.
The Coulomb energy
\begin{equation}
  \label{eq:charging}
  E_c= \frac{e^2}{4\pi \varepsilon_{0}\kappa d},
\end{equation}
where $\kappa$ is the effective dielectric constant of the NC array is another important energy scale.
For all NCs with $d<8 ~\mathrm{nm}$ one finds  $\Delta_{e,h} > E_{c}$, so we concentrate on this case.

In this paper we deal with epitaxially connected NCs and propose  a  mechanism where the electron and hole tunnel  through the small $\rho$ contact facet in tandem. This happens in the second order perturbation theory  through the intermediate  state with energy $\xi E_{c}$ in which the electron is already in the right NC, while the hole is still  in the left NC. We show below that $\xi$ is very close to unity. The matrix element of the tandem tunneling is

\begin{equation}
  \label{eq:M_T}
M_T= 2 \frac{t_e t_h}{E_c},
\end{equation}
where 
\begin{equation}
  \label{eq:t_2}
t_{e,h}=\frac{8}{3\pi} \Delta_{e,h} \left(\frac{\rho}{d}\right)^{3}  
\end{equation}
are matrix elements for the electron and hole single particle  tunneling through epitaxial connection from one NC to another, respectively. Eq. (\ref{eq:t_2}) was implicitly derived in Ref. \cite{localization_length_NC} and because of important role of $t_{e,h}$ in our theory we repeat its derivation below.

Let us formulate conditions at which our theory is applicable.
We consider the case when an exciton  does not dissociate.
An exciton ionization requires energy larger than $E_{c}$ because two distant charged NCs are created from the neutral background.
Thus,  we consider  low temperatures $k_{B} T \ll E_c$, when  excitons are not thermally ionized.
Diameters  of a NC typically vary by $\alpha d$, where $\alpha \simeq 0.05-0.1$ \cite{Murray-AnnuRev30-2000}.
This leads to variation of the electron energy by $2 \alpha \Delta$.
We  assume that $E_c \gg 2 \alpha \Delta$, so that  electron and hole cannot  move separately.
We also assume that  $\rho/d < \alpha^{1/3}$, so that  $t_{e,h} \ll  2 \alpha \Delta_{e,h}$.
In this situation  electrons, holes and excitons  are localized in their NCs.
The opposite case $t_{e,h} \gg 2 \alpha \Delta_{e,h}$, when electrons and holes are delocalized in the NC array  was studied in Refs. \cite{Delerue_band,Delerue_2D_band,Delerue_topology_2D}.

Thus, we  deal with the situation where
\begin{equation}
  \label{eq:condition}
t\ll 2\alpha \Delta \ll E_{c} \ll \Delta.
\end{equation}
Energies $t,\alpha\Delta,E_{c},\Delta$ are estimated for the wide class of NC materials in Table 1, where we see that all our conditions Eq. (\ref{eq:condition}) are fulfilled (although sometimes only marginally).
As shown in Table 1 for majority of materials   the tandem tunneling exciton transfer rate is larger than the Dexter rate. For Si it is even much larger than F{\"o}rster rate.

\begin{table}[t!]
\label{tab:parameters}
  \begin{tabular}{ c | c| c|  c|  c|  c |c|  c|  c|c}
    \hline
    NC & $\sqrt{m_{e}m_{h}}/m$  & $\kappa_{NC}$  & $a_{0} ~~\mathrm{\AA}$ &$ \Delta_{e}$&$2 \alpha \Delta_{e}$&$E_{c}$ &$t_{e}$&$\tau_{T}^{-1}/\tau_{F}^{-1}$ & $\tau_{T}^{-1}/\tau_{D}^{-1}$  \\ \hline
    InP & 0.1 &9.6 & 7 &490&50&60&3& 0.2 & $10^{2}$  \\
    CdSe & 0.16 &9.5 & 5 &400&40&60&3& 0.1 & $20$  \\ 
    ZnO & 0.26 &3.7& 2&140&15&150&1&$10^{-2}$ & 0.05  \\ 
    Si & 0.22 & 12 & 0.4 &150&15&50&1&$10^3$ & 10 \\
    \hline
  \end{tabular}
  \caption{Parameters and results for  different NCs. Effective  mass $\sqrt{m_{e}m_{h}}$ is in the units of electron mass $m$, $\kappa_{NC}$ is the high frequency dielectric constant of the material, $\alpha=0.05$. Charging energy $E_{c}$, ground state kinetic energy  $\Delta_{e}$ and matrix  element for the electron tunneling $t_{e}$ are in $~\mathrm{meV}$.  Ratios of tandem tunneling $\tau_{T}^{-1}$ to F{\"o}rster $\tau_{F}^{-1}$ or Dexter $\tau_{D}^{-1}$ rates are estimated  for the case $d = 6~\mathrm{nm}$ and $\rho = 0.2 d  \simeq 1~\mathrm{nm}$. $ea_{0}$ is the dipole moment of the interband transition}
\end{table}

\section{Results and Discussion}
Let us first  formulate our results.
We show that in epitaxially connected array of NCs, the ratio between the tandem tunneling $\tau_{T}^{-1}$ and F{\"o}rster rates $\tau_{F}^{-1}$  is
\begin{equation}
  \label{eq:the_answer_foster}
  \frac{\tau_{T}^{-1}}{\tau_{F}^{-1}} = \left(8.7 \frac{a_{B}}{a_{0}}\right)^{4} \left(\frac{\kappa_{NC}+2\kappa}{\kappa_{NC}} \right)^{4} \left(\frac{\rho}{d}\right)^{12}.
\end{equation}
Here  $a_{B} = 4\pi \hbar^{2} \varepsilon_{0} \kappa_{NC}/\sqrt{m_{e}m_{h}} e^{2}$ is the unconventional effective exciton  Bohr radius, $e a_{0}$ is the dipole moment matrix element taken between the valence- and conduction-band states and $\kappa_{NC}$ is the high frequency dielectric constant of the material. 

In the Table 1 we summarize our estimates of the ratio (\ref{eq:the_answer_foster}) for different NCs. We used  $d = 6~\mathrm{nm}$ and $\rho= 0.2 d \simeq 1 ~\mathrm{nm}$. Values of $\kappa_{NC}$ are taken from Ref. \cite{978-3-642-62332-5}. For epitaxially connected NCs we use  $\kappa=2\kappa_{NC}\rho/d$  (see  Ref. \cite{dielectric_constant_touching_NC}).

The ratio (\ref{eq:the_answer_foster}) is derived for materials with an isotropic single band hole $m_{h}$ and electron $m_{e}$ masses. For most materials the spectra are more complex. Below we explain how we average the masses for these materials and also how we calculate $a_{0}$.

We see that, the tandem tunneling can be comparable with the F{\"o}rster mechanism in semiconductors  like InP, CdSe where the effective mass is small.
The tandem tunneling can be more efficient in cases where the F{\"o}rster mechanism is forbidden.
For example, in indirect band gap semiconductors like Si, where  $a_{0}$ is small and  the F{\"o}rster mechanism is not effective, the tandem tunneling mechanism dominates.

In another situation the tandem tunneling dominates  at low temperatures. Excitons can be in bright or dark spin states \cite{Efros_dark_exciton}. Only the bright exciton can hop due to the F{\"o}rster mechanism. The dark exciton has smaller energy and the  dark-bright exciton splitting is of the order of a few meV. So at small temperatures an exciton is in the dark state and cannot hop by the F{\"o}rster mechanism. At the same time the tandem tunneling is not affected by a spin state of an exciton.

Dexter \cite{Dexter_transfer} suggested another exciton transfer mechanism which also is not affected by spin state of an exciton. Two electrons of two NCs exchange with each other (see Fig. \ref{fig:Scheme}c). We show below that for an array of NCs the ratio between rates for tandem tunneling and the Dexter mechanism is:
\begin{equation}
  \label{eq:the_answer}
  \frac{\tau_{T}^{-1}}{\tau_{D}^{-1}} = \left(\frac{\Delta_{e}\Delta_{h}}{4E_{c}^{2}}\right)^{2}.
\end{equation}
In most cases $\Delta_{e,h} \gg E_{c}$ and as one can see from  Table 1 that the tandem tunneling rate  is much larger than the Dexter rates with the exception  of ZnO.

It is worth noting that the same ratio holds not only for epitaxially connected  NCs but for NCs separated by ligands.
Of course, if NCs are separated by ligands say by distance $s$ and wave functions decay in ligands as $\exp(-s/b)$, where $b$ is the decay length of an electron outside of a NC, both rates acquire additional factor $\exp(-4s/b)$.
Also,  the difference between the tandem mechanism and Dexter transfer emerges only in NCs, where  $\Delta_{e,h} \gg E_{c}$.
In atoms and molecules, where essentially $E_{c} \simeq \Delta$  there is no such difference between the two mechanisms.

For epitaxially connected Si and InP NCs where the tandem tunneling  is substantial these predictions can be verified in the following way.
One can transform the bright exciton to the dark one by  varying  magnetic field or temperature.
The exciton in the dark state cannot hop by the F{\"o}rster mechanism, and usually hops much slower \cite{energy_transport_NC_Rodina,Exciton_CdSe_H_T}.
For epitaxially connected NCs, where the tandem  rate is larger than the F{\"o}rster one the exciton transfer should not be affected  by magnetic field or temperature.

Let us switch to derivation of the main result. For that we first should discuss electron wave functions in epitaxially connected NCs.

\emph{Wave functions of two epitaxially connected NCs.} Below we describe the envelope wave functions  in two epitaxially connected NCs. Here we present only scaling estimates  and calculate  numerical coefficients in the methods section.  The  wave functions for electrons and holes are the same, so we concentrate only on the electron.
In  an isolated NC the electron wave function is:

\begin{equation}
  \label{eq:psi_0}
  \psi_{0}(r) = \frac{1}{\sqrt{\pi d} r} \sin \left(2\pi \frac{r}{d}\right),
\end{equation}
where $r$ is the distance from the center of the NC. We focus on two NCs shown on Fig \ref{fig:Scheme}, which touch each other by the small facet in the plane $z=0$. In this situation the wave function for an electron  in the left NC $\Psi^{L}$ leaks through this small facet, so that it is finite  in the plane of the facet $z=0$ and in the right NC.  The derivative $\partial \Psi^{L}/\partial r$ is hardly changed by this small perturbation, so that the wave function  in the plane $z=0$  acquires a finite value:
\begin{equation}
  \label{eq:4}
  \Psi^{L}(z=0) \simeq \rho \frac{\partial \psi_{0}}{\partial z} \simeq \frac{\rho}{d^{5/2}}.
\end{equation}

The same happens with the wave functions of an electron  in the right NC $\Psi^{R}$. $\Psi^{L}$ and $\Psi^{R}$ are symmetric with respect to the plane $z=0$.

\emph{Tunneling matrix element.} We calculate the matrix element (\ref{eq:M_T}) of an electron and hole tunneling  through the contact facet in the second order perturbation theory.
$E_c$ is the energy of the intermediate state, in which the electron moves to the right NC, while the hole is still in the left NC.
In other words the left NC plays the role of donor (D) and the right one the role of acceptor (A) so that intermediate state is $D^{+}A^{-}$ state.
For touching NCs the energy of $D^{+}A^{-}$ state is evaluated in the methods section and is shown to be $\xi E_{c}$, where $|\xi-1| < 0.1$.
Therefore in Eq. (\ref{eq:M_T}) and through out the paper we use $\xi=1$.
In Eq. (\ref{eq:M_T}) factor $2$  takes care about two possible orders of electron and hole hops.

Matrix elements $t_e,t_h$ for the electron and hole single particle  tunneling from one NC to another can be written as \cite{landau_quantum} (see the methods section)   
\begin{equation}
  \label{eq:t}
  t_{e,h}=\frac{\hbar^2}{m_{e,h}} \int \Psi^{L*} (r_1) \frac{\partial}{\partial z} \Psi^{L} (r_1) dS,
\end{equation}
where the integration is over the plane $z=0$.
Using Eqs. \eqref{eq:psi_0},~(\ref{eq:4})  we arrive to (\ref{eq:t_2}).
Substituting (\ref{eq:t_2}) into Eq. (\ref{eq:M_T}) we get
\begin{equation}
  \label{eq:our_Dexter}
M_{T}= C_{T} \frac{\Delta_{e}\Delta_{h}}{E_{c}}\left(\frac{\rho}{d}\right)^{6},
\end{equation}
where  the numerical coefficient $C_{T}=2^{7}/9\pi^{2} \simeq 1.4$ is calculated in the methods section.

Above we assumed that the energy spectra of electrons and holes are isotropic and have one band. In fact in most cases the hole energy spectrum has heavy and light band branches with masses $m_{hh}$ and $m_{hl}$ respectively. The energy of the lower state $\Delta_{h}$ can be determined with adequate accuracy if instead of a complicated valence band structure we consider  a simple band ~\cite{Yassievich_excitons_Si,Holes_electrons_NCs} in which the holes have an average  mass  $m_{h}=3m_{hl}m_{hh}/(m_{hl}+2 m_{hh})$. For indirect band materials like Si an electron in the conduction band has an anisotropic  mass in transverse $m_{et}$ and parallel  $m_{ep}$ directions. The effective mass $m_{e}$, which determines the energy of the lower state $\Delta_{e}$ has a similar form $m_{e}=3m_{et}m_{ep}/(m_{et}+2 m_{ep})$. Using data for the electron and hole masses from Ref. \cite{978-3-642-62332-5} we get the values $\sqrt{m_{e}m_{h}}$ which is shown in the Table 1.

\emph{F{\"o}rster matrix element.} Now we dwell  on the F{\"o}rster matrix element. It is known \cite{Delerue_Foster_NC} that the matrix element for  the F{\"o}rster transfer between two touching NCs  is

\begin{align}
  \label{eq:M_F}
  M_F & = \sqrt{\frac{2}{3}} \frac{e^{2}}{4\pi \varepsilon_{0} d^{3}} \eta a_{0}^{2}.
\end{align}
Here we assume that dipoles which interact with each other are concentrated in the center of NCs. The factor $\eta=9\kappa/(\kappa_{NC}+2\kappa)^{2}$ takes into account that the dipole-dipole interaction is screened \cite{Forster_Rodina}. The product $e a_{0}$ is the matrix element of the dipole moment between the conduction and valence band. Eqs.  (\ref{eq:our_Dexter}) and (\ref{eq:M_F}) bring us to the ratio  (\ref{eq:the_answer_foster}).

In order to find $a_{0}$ we note that the matrix element of dipole moment is related to the band gap $E_{g}$ of a material and the  momentum matrix element $p$ as ~\cite{laser_devices_Blood} 

$$a_{0}^{2} = \frac{\hbar^{4}p^{2}}{m^{2}E_{g}^{2}}.$$

According to the Kane model $p$ determines the effective electron mass \cite{Efros_review_NCs}, so we can say that

\begin{equation}
  \label{eq:a0}
a_{0}^{2}=\frac{3}{4} \frac{\hbar^{2}}{E_{g}m_{e}}.  
\end{equation}
The estimate for $a_{0}$ for direct gap materials is given in the Table 1. 
For an indirect band gap semiconductor such as Si the dipole-dipole transition is forbidden. However, in small  NCs this transition is possible due to the confinement or the phonon assistance.  One can get estimate of the effective  $a_{0}$ in the following way. The transfer rate for InAs is $10^{7}$ times larger than for Si \cite{Delerue_Foster_NC}, because their dielectric constants are close we assume that the difference in rates is due to $a_{0}$. Thus for Si, effective $a_{0}$ is $55$ times smaller than for InAs, which we get with the help of the Eq. (\ref{eq:a0}).

\emph{Dexter matrix element.} The physics of the Dexter transfer mechanism\cite{Dexter_transfer} involves electron tunneling, but differs from that of the tandem tunneling mechanism in the following sense.
The Dexter matrix element $M_{D}$ is calculated below  in the first order perturbation theory in electron-electron interaction between two-electron wave function.
The tandem tunneling matrix element was  calculated  in Eq. (\ref{eq:M_T}) in the second order perturbation theory, where $t_e$ and $t_h$ are single particle transfer integrals  calculated between one-electron wave functions.
Here we calculate the  Dexter matrix element and show that at $\Delta \gg E_{c}$  it is much  smaller than the tandem one. It is easier to consider this mechanism in the electron representation.
The Dexter exciton transfer happens due to potential exchange interaction between two electrons in NCs.
The initial state is $\Psi^{L*} (r_1) \Psi^{R}(r_{2})$ \textit{i.e.} the first electron in the conduction band of the left NC and the second electron is in the valence band of the right NC.
The final state is $\Psi^{R} (r_1) \Psi^{L} (r_2)$, \textit{i.e.} the  first electron in the conduction band of the right NC and the second electron in the valence band of the left NC (see Fig. \ref{fig:Scheme} a).
The matrix element has the following form:
\begin{equation}
   \label{eq:M_D}
   M_D = \int \Psi^{L*} (r_1) \Psi^{R*} (r_2) V(r_{1},r_{2}) \Psi^{R} (r_1) \Psi^{L} (r_2) d^3r_1 d^3 r_2.
\end{equation}
Here $V(r_{1},r_{2})$ is the  interaction energy between electrons in points $r_{1}$ and $r_{2}$, which is of the order of $E_{c}$. In general,  calculating the matrix element is a difficult problem.  For our case, however, a significant simplification is available because the internal dielectric constant $\kappa_{NC}$ is typically  much larger than  the external dielectric constant $\kappa$ of the insulator in which the NC is embedded. The large internal dielectric constant $\kappa_{NC}$ implies that the NC charge is  homogeneously redistributed over the NC surface.  As a result a semiconductor NC  can be approximately considered as a metallic one in terms of its Coulomb interactions, namely that when  electrons are in two different NCs, the NCs are neutral and there is no interaction between them and $V=0$. When electrons are in the same NC, both  NCs are charged and  $V=E_{c}$. Thus, we can approximate Eq. (\ref{eq:M_D}) as:
\begin{equation}
  \label{eq:M_D_2}
  M_D = 2 E_{c} \left(\int \Psi^{L}(r)  \Psi^{R}(r) d^{3}r\right)^{2}.
\end{equation}
The integral above is equal to  $2 t_{e}/\Delta_{e}$ (see methods section) and we get:

\begin{equation}
  \label{eq:7}
M_D =  C_{D} E_{c}\left(\frac{\rho}{d}\right)^{6},
\end{equation}
where  $C_{D} = 2^{9}/9\pi^{2} \simeq 5.7$ is the numerical coefficient.
Let us compare Eqs. (\ref{eq:7}) and (\ref{eq:our_Dexter}) for matrix elements $M_D$ and $M_T$ of Dexter and tundem processes.
We see that $M_D$ is proportional to $E_c$, while $M_T$ is inverse proportional to $E_c$.
(The origin of this difference is related to the fact that in Anderson terminology \cite{Anderson_potential_kinetic_exchange} the former one describes ``potential exchange'', while the latter one describes ``kinetic exchange''.
In the magnetism theory \cite{Anderson_potential_kinetic_exchange} the former leads to ferromagnetism and the  latter to antiferromagnetism).
Note that the ratio (\ref{eq:the_answer}) is inverse proportional to the fourth power of the effective mass.
As a result in semiconductors with small effective mass such as  InP and CdSe the ratio of tandem and Dexter rates is very large (up to $100$).
Using Ref. \cite{dielectric_constant_touching_NC}  $\kappa=2\kappa_{NC}\rho/d$ in the Table 1 we calculate the ratio for different NCs.
We see that typically  the tandem tunneling rate is larger or comparable with the Dexter one.

So far we dealt only with NCs in which the quantization energy $\Delta$ is smaller than  half of the  semiconductor energy gap and one can use  parabolic electron and hole spectra. This condition is violated in semiconductor NCs with very small effective masses $\sim 0.1 ~m$ and small energy gaps $\sim 0.2 \div 0.3$ eV such as InAs and PbSe. In these cases, the quantization energy  $\Delta$ should be calculated using non-parabolic ("relativistic") linear part of the electron and hole spectra $|\epsilon| = \hbar v k$, where $v \simeq 10^{8} ~\mathrm{cm/s}$~\cite{Wise_PbSe,PbS_spectrum,PbSe_NC_spectrum_Delerue}. This gives $\Delta = 2 \pi \hbar v/d$. We show in the methods section that substitution of $\Delta_{e,h}$  in Eq. (\ref{eq:t_2}) by $\Delta/2$  leads to the correct ``relativistic'' modification of the single particle tunneling matrix element $t$ between two such NCs. Then for InAs and  PbSe NCs with the same geometrical  parameters as in the Table 1 we arrive at  ratios $\tau^{-1}_{T}/\tau^{-1}_{D}$ as large as $1000$ (see Table 2). One can see however that inequalities (\ref{eq:condition}) are only  marginally valid so that this case deserves further attention.

\begin{table}[h]
\label{tab:parameters}
  \begin{tabular}{ c | c| c|  c|  c|  c |c|  c |c }
    \hline
    NC & $\kappa_{NC}$  & $a_{0} ~~\mathrm{\AA}$ &$ \Delta$&$\alpha \Delta$&$E_{c}$ &$t_{e}$&$\tau_{T}^{-1}/\tau_{F}^{-1}$ & $\tau_{T}^{-1}/\tau_{D}^{-1}$  \\ \hline
    PbSe & 23 &25.8 &660 &33&25&2&0.1& $10^{3}$  \\
    InAs & 12.3 &19.7 &660 &33&46&2&$10^{-2}$& $10^{2}$  \\
    \hline
  \end{tabular}
  \caption{Parameters and results for ``relativistic'' NCs PbSe and InAs. As in the  Table 1 we use $d=6 ~\mathrm{nm}$, $\rho=0.2 d \simeq 1 ~\mathrm{nm}$ and $\alpha=0.05$}
\end{table}

\section{Conclusion}
In this paper,  we considered the exciton transfer in the array of epitaxially connected through the facets with small radius $\rho$ NCs.
After evaluation of matrix elements for  F{\"o}rster and Dexter rates  in such arrays we proposed an alternative mechanism of tunneling of the exciton where electron and hole tunnel in tandem  through the contact facet. The tandem tunneling happens in the second order perturbation theory  through the intermediate  state in which the electron and the hole are in different NCs.
For all semiconductor NCs we studied except ZnO the tandem tunneling rate  is much larger than the Dexter one.
The tandem tunneling rate is comparable with  the F{\"o}rster one for bright excitons and dominates for dark excitons.
Therefore it determines exciton transfer at low temperatures.
For silicon NCs the tandem tunneling rate  substantially  exceeds the F{\"o}rster rate.

\section{Methods}

\subsection{Calculation of $M_{T}$}
\label{sec:Appendix_wave_function_t}

If two NCs are separated their 1S ground state is degenerate. When they touch each other by small facet with radius $\rho \ll d$, the degeneracy is lifted and the 1S state is split into two levels $U_{s}$ and $U_{a}$ corresponding to the electron wave functions:

\begin{equation}
  \label{eq:psi_g_u}
  \psi_{s,a}=\frac{1}{\sqrt{2}} [ \Psi^{L}(-z) \pm \Psi^{L}(z)],
\end{equation}
which are symmetric and antisymmetric about the plane $z=0$. The difference between two energies $U_{a}-U_{s}=2 t$, where $t$ is the overlap integral between NCs. Similarly to  the problem 3 in  $\S$50 of Ref. \cite{landau_quantum} we get Eq. (\ref{eq:t}).

Below we find $\Psi^{L}$  in the way which is outlined in \cite{Rayleigh_diffraction_aperure,localization_length_NC}. We look for solution in the form

\begin{equation}
  \label{eq:psi_definition}
  \Psi^{L}=\psi_{0}+\psi,   
\end{equation}
where $\psi_{0}$ is non-zero only inside a NC. $\psi$ is the correction which is substantial only near the contact facet with the radius $\rho \ll d$ so  $\nabla^{2} \psi \gg \psi \Delta_{e,h}$ and we can omit the energy term in the  Schrodinger equation:

\begin{equation}
  \label{eq:Laplacian}
  \nabla^{2} \psi=0.
\end{equation}
Near the contact facet with $\rho \ll d$ two touching spheres can be seen as an impenetrable plane screen and the the contact facet as the aperture in the screen. The boundary conditions for $\psi$ are the following: $\psi=0$ on the screen, while in the aperture the derivative $d\Psi^{L}/dz$ is continuous:

\begin{equation}
  \label{eq:psi_derivative}
  \left . \frac{\partial \psi}{\partial z} \right|_{z=+0} =    \left . \frac{\partial \psi_{0}}{\partial z} + \frac{\partial \psi}{\partial z} \right|_{z=-0}.
\end{equation}

As shown in Refs. \cite{Rayleigh_diffraction_aperure, localization_length_NC} $\psi$ is symmetric with respect to the plane $z=0$.  As a result,
\begin{equation}
  \label{eq:psi_z}
  \left. \frac{\partial \psi}{\partial z} \right|_{z=+0} = -2\frac{\sqrt{\pi}}{d^{5/2}} =A.
\end{equation}

\begin{figure}[ht]
  \includegraphics[width=0.45\textwidth]{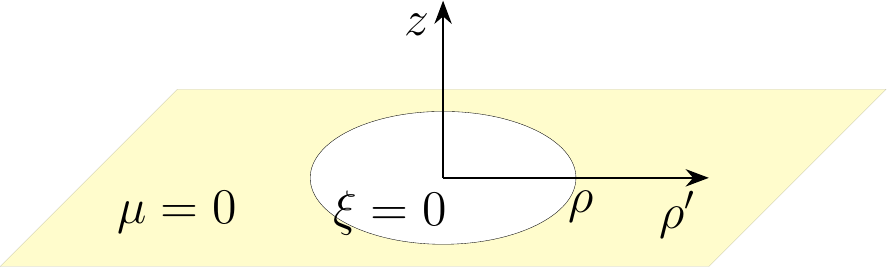}
  \caption{\label{fig:oblate} The contact with radius $\rho$ between two spheres with diameter $d \gg \rho$ can be  represented by a screen with an aperture. In oblate spheroidal coordinates the aperture corresponds to the plane $\xi=0$ and the screen corresponds to the plane $\mu=0$ }
\end{figure}

It is easy to solve the Laplace equation with such boundary condition in the oblate spheroidal coordinates $\varphi$ ,$\xi$, $\mu$, which are related with  cylindrical coordinates $z$, $\rho'$, $\theta$  (see Fig.~\ref{fig:oblate}) as

\begin{eqnarray}
  \rho' & = &\rho \sqrt{(1+\xi^{2})(1-\mu^{2})} \nonumber \\
  z & = &\rho \xi \mu  \\
  \varphi & =& \varphi \nonumber 
\label{eq:relation}
\end{eqnarray}

The Laplace  equation can then be rewritten \cite{9780387184302}:

\begin{equation}
  \label{eq:Laplace}
  \frac{\partial}{\partial \xi} (1+\xi^{2}) \frac{\partial \psi}{\partial \xi} +   \frac{\partial}{\partial \mu} (1-\mu^{2}) \frac{\partial \psi}{\partial \mu}=0.
\end{equation}
The boundary conditions in this coordinates will be $\psi=0$ for $\mu=0$ ($z=0$ and $\rho'>\rho$) and for the region $\xi=0$ ($z=0$, $\rho'<\rho$)

$$
\left. \frac{\partial \psi}{\partial \xi}\right|_{\xi=0} \frac{1}{\rho \mu} = A.
$$

One can check by direct substitution that the solution at $z>0$ of the equation with these boundary conditions is:

\begin{equation}
  \label{eq:solution}
  \psi=  \frac{2\rho A}{\pi} \mu (1-\xi \arccot \xi)
\end{equation}

Thus in the contact between two spheres $\xi=0$ ($z=0$, $\rho'<\rho$):

\begin{equation}
  \label{eq:surface_psi}
  \psi= \frac{4}{\sqrt{\pi}} \frac{1}{d^{3/2}} \frac{\rho}{d} \sqrt{1-\frac{\rho'^{2}}{\rho^{2}}}.
\end{equation}

Now we can calculate the integral (\ref{eq:t}) using expression for $\psi$ in the contact between two NCs (\ref{eq:surface_psi}) and arrive at Eq. (\ref{eq:t_2}).

\subsection{The energy of the intermediate state}
Here we study a cubic lattice of touching NCs with the period $d$.
For large $\kappa_{NC}$ it can be considered as the lattice of identical capacitors with capacitance $C_{0}$ connecting nearest neighbor sites.
One can immediately get that the macroscopic dielectric constant of the NC array is  $\kappa=4\pi C_{0}/d$. 
We calculate the energy for the intermediate state, where an electron  and a hole occupy the nearest-neighbor NC, the reference point of energy being energy of all neutral NCs.
The Coulomb energy necessary to add one electron (or hole) to a neutral NC is called the charging energy $E_{e}$ .
It was shown \cite{dielectric_constant_touching_NC} that for touching NCs which are arranged in the cubic lattice this energy is:
\begin{equation}
  \label{eq:chargin_energy}
  E_{e}=1.59 E_{c}.
\end{equation}
We show here that the interaction energy between two nearest neighbors NC is $ E_{I}=-2\pi/3 E_{c}$, so that the energy of the intermediate state is $2E_{e}+E_{I} =\xi E_{c}$, where $\xi \simeq 1.08$.
Let us first remind the derivation of the result (\ref{eq:chargin_energy}).

By the definition the charging energy is
\begin{equation}
  \label{eq:charging_energy_capacitance}
  E_{e} = \frac{e^{2}}{2C},
\end{equation}
where $C$ is the capacitance of a NC immersed in the array.
It is known that the capacitance between a site in the cubic lattice  made of identical capacitance $C_{0}$ and the infinity is $C=C_{0}/\beta$, $\beta \simeq 0.253$ ~\cite{PhysRevB.70.115317,Lattice_green_function}.
We see that $1/\beta$ plays the role of the effective number of parallel capacitors connecting this site to infinity. Thus we arrive at
\begin{equation}
  \label{eq:charging_energy_capacitance_2}
  E_{e} = \frac{e^{2}}{2C} =2\pi \beta E_{c} \simeq 1.59  E_{c}
\end{equation}
Here we also need the interaction energy between two oppositely charged nearest sites of the cubic lattice.
\begin{equation}
  \label{eq:ineraction_energy}
  E_{I} = -\frac{e^{2}}{2 C_{12}},
\end{equation}
where $C_{12}$ is the total capacitance between the two nearest-neighbor NCs. It is easy to get that $C_{12}=3C_{0}$, so that 

\begin{equation}
  \label{eq:ineraction_energy}
  E_{I} = -\frac{e^{2}}{2 C_{12}} = -\frac{2\pi}{3}E_{c}.
\end{equation}
Thus we arrive at the energy of the intermediate state for the cubic lattice: $2E_{e}+E_{I} \simeq 1.08 E_{c}$, \textit{i.e.} for this case we get $\xi=1.08$.
We repeated this derivation for other lattices. We arrived at  $\xi=0.96$ and $\xi=0.94$ for bcc and fcc latices of capacitors, respectively.

\subsection{Calculation of $M_{D}$}
\label{sec:Dexter}
One can calculate the integral (\ref{eq:M_D_2}) in the following way. $\Psi^{R}$ in the left NC can be written as $\psi$. We start from the second Green identity for functions $\Psi^{L}$ and $\psi$:

\begin{equation}
  \label{eq:Green_identity}
  \int d^{3}r (\psi \nabla^{2} \Psi^{L} - \Psi^{L} \nabla^{2} \psi ) = \int dS (\psi \nabla \Psi^{L} - \Psi^{L} \nabla \psi),
\end{equation}
Because $\psi$  satisfies the Eq. (\ref{eq:Laplacian}) and $\Psi^{L}$ is zero on the surface of a NC except the contact facet, where it is equal to $\psi$ we get:

$$  \int \psi(r) \Psi^{L}(r)  d^{3}r  = \frac{2t}{\Delta} $$

\subsection{Non-parabolic band approximation}
Below we use non-parabolic ``relativistic'' Kane approach ~\cite{PbS_spectrum}. Namely we assume that the wave function $\psi_{0}$ of an electron and hole in the ground state of the isolated spherical NC satisfies Klein-Gordon equation:
\begin{equation}
  \label{eq:KG}
  -\hbar^{2}v^{2} \Delta \psi_{0}+m^{*2}v^{4}\psi_{0}=E^{2}\psi_{0}.
\end{equation}
This approximation works well for the ground state of an electron and hole~\cite{PbS_spectrum}. The energy spectrum is:
\begin{equation}
  \label{eq:spectrum}
  E(k)=\pm\sqrt{m^{*2}v^{4}+\hbar^{2}v^{2}k^{2}}.
\end{equation}
One can immediately see that the bulk band gap $E_{g}=2m^{*}v^{2}$. The solution of the equation (\ref{eq:KG}) for spherical isolated NC is the same as in the parabolic band approximation (see  Eq. (\ref{eq:psi_0})). The kinetic energy $\Delta$ becomes:

\begin{equation}
  \label{eq:Delta_relativistic}
  \Delta=\sqrt{m^{*2}v^{4}+\hbar^{2}v^{2}\left(\frac{2\pi}{d}\right)^{2}}-m^{*}v^{2}.
\end{equation}

Let us now concetrate on the expression for $t$. If two NCs are separated their 1S ground state is degenerate. When they touch each other by small facet with radius $\rho \ll d$, the degeneracy is lifted and the 1S state is split into two levels $U_{s}$ and $U_{a}$ corresponding to the electron wave functions:

\begin{equation}
  \label{eq:psi_g_u}
  \psi_{s,a}=\frac{1}{\sqrt{2}} [ \Psi^{L}(-z) \pm \Psi^{L}(z)],
\end{equation}
which are symmetric and antisymmetric about the plane $z=0$. The difference between two energies $U_{a}-U_{s}=2 t$, where $t$ is the overlap integral between NCs. Similarly to  the problem 3 in  $\S$50 of Ref. \cite{landau_quantum} we use that $\psi_{s}$ satisfies the Eq. (\ref{eq:KG}) with the energy $U_{s}$ and $\Psi^{L}$ satisfies the same equation with the energy $E_{L}$. As a result we get the difference:

\begin{equation}
  \label{eq:difference}
  E^{2}_{L}-U^{2}_{s} = \hbar^{2} c^{2} \int \left(\Psi_{L}\Delta \psi_{s} - \psi_{s} \Delta \Psi_{L} \right) dV \left( \int \psi_{s} \Psi_{L}  dV \right)^{-1}
\end{equation}

Repeating the same step for $\psi_{a}$ we arrive at:

\begin{equation}
  \label{eq:t_resistivity}
  t=(U_{a}^{2}-U_{s}^{2}) /4 =\frac{\hbar^{2}v^{2}}{U_{s}} \int \Psi^{L} \frac{\partial \Psi^{L}}{\partial z} dS.
\end{equation}
One can check that this expression at $m^{*}v^{2} \gg \hbar v k$ leads to (\ref{eq:t}).
For $m*=0$ we get:

\begin{equation}
  \label{eq:t_r}
  t = \frac{\hbar v d}{2\pi} \int \Psi^{L} \frac{\partial \Psi^{L}}{\partial z} dS.
\end{equation}
Using the same approach for the calculation of the integral as in S1 we get:

\begin{equation}
  \label{eq:t_R}
  t=\frac{4}{3\pi} \Delta \left(\frac{\rho}{d}\right)^{3}
\end{equation}

In that case the Eq. (\ref{eq:the_answer_foster}) for the ratio tandem and F{\"o}rster rates can be written as:

\begin{equation}
  \label{eq:Tandem_forster}
  \frac{\tau_{T}^{-1}}{\tau_{F}^{-1}}=3.7 \left(\frac{\hbar v}{e^{2}}\right)^{4} \left(\frac{d}{a_{0}}\right)^{4} (\kappa+2\kappa_{NC})^{4} \left(\frac{\rho}{d}\right)^{12}.
\end{equation}

and the ratio between tandem and Dexter rates is:

\begin{equation}
  \label{eq:Tandem_dexter}
  \frac{\tau_{T}^{-1}}{\tau_{D}^{-1}}=\frac{\pi^{4}}{2^{4}} \left(\frac{\hbar v}{e^{2}}\right)^{4}  \kappa_{NC}^{4}.
\end{equation}

Eqs. (\ref{eq:Tandem_forster}) and (\ref{eq:Tandem_dexter}) are used to calculate the ratio in the Table 2.

\section{Acknowledgement}
We are grateful to A. V. Chubukov, P. Crowell, Al. L. Efros, H. Fu, R. Holmes, A. Kamenev, U. R. Kortshagen, A. V. Rodina,  I. Rousochatzakis,  M. Sammon, B. Skinner, M.V. Voloshin, D. R. Yakovlev and I. N. Yassievich  for helpful discussions. This work was supported primarily by the National Science Foundation through the University of Minnesota MRSEC under Award No. DMR-1420013.

\providecommand{\latin}[1]{#1}
\providecommand*\mcitethebibliography{\thebibliography}
\csname @ifundefined\endcsname{endmcitethebibliography}
{\let\endmcitethebibliography\endthebibliography}{}

\end{document}